\documentclass[12pt]{article}
\usepackage{graphicx}
\usepackage{cite}

\date{}
\begin{document}

\title{Two-dimensional nonlocal multisolitons}

\maketitle
\begin{center}
\author{V.M. Lashkin$^1$, A.I. Yakimenko$^{1,2}$, and O.O.
Prikhodko$^{2}$}

\small{$^1$Plasma Theory Department, Institute for Nuclear
Research, Kiev 03680, Ukraine}\\
\small{$^2$Department of Physics, Kiev University, prosp.
Glushkova 6, Kiev 03022, Ukraine}
\end{center}

\begin{abstract}
We study the bound states of two-dimensional bright solitons in
nonlocal nonlinear media. The general properties and stability of
these multisolitary structures are investigated analytically and
numerically. We have found that a steady bound state of coherent
nonrotating and rotating solitary structures (azimuthons) can
exist above some threshold power. A dipolar nonrotating
multisoliton occurs to be stable within the finite range of the
beam power. Azimuthons turn out to be stable if the beam power
exceeds some threshold value. The bound states of three or four
nonrotating solitons appear to be unstable.
\end{abstract}


The recent experimental observations of spatial solitons in
nonlocal media such as nematic liquid crystals
\cite{ContiPRL03,ContiPRL04}, lead glasses \cite{Segev}, renewed
an interest to coherent structures in spatially nonlocal nonlinear
media. In the spatially nonlocal media the nonlinear response
depends on the wave packet intensity at some extensive spatial
domain. Nonlocality is a key feature of many nonlinear media. It
naturally appears in different physical systems such as plasmas
\cite{Litvak1,LitvakSJPlPhys75,Davydova}, Bose- Einstein
condensates \cite{Pedri}, optical media \cite{Krolik},  liquid
crystals \cite{ContiPRL03}, and soft matter \cite{Trillo}.
Different types of single (solitons , vortices) and composite
(soliton clusters) solitary structures have been predicted and
experimentally observed in various nonlinear media
\cite{Kivshar1}. Rotating scalar multipoles (azimuthons) were
first introduced in Ref. \cite{Kivshar2} for models with local
nonlinearity.

A composite soliton structure, or a multi-soliton complex is a
self-localized state which is a nonlinear superposition of several
fundamental solitons. Stationary multi-soliton structures in
nonlocal media were considered first in Refs.
\cite{Mironov1,Mironov2}. However, a question of the stability of
the multisolitons is still an open problem. In particular, the
form of the nonlinear nonlocal response is crucial for the
multisoliton stability. Stable two-dimensional (2D) rotating
\cite{Kivshar3} and nonrotating \cite{We} dipole solitons in a
medium with Gaussian response function have been numerically
observed. Authors of Refs. \cite{vector} investigated a more
realistic model with the Helmholtz type response function (often
referred to as the model with thermal nonlinearity) and showed
that multipole vector solitons can be made stable. This model
describes, in particular, the nonlinear nonlocal response of some
thermo-optical materials, liquid crystals, and partially ionized
plasmas \cite{ContiPRL03,Segev,LitvakSJPlPhys75}. Very recently
 experimental observation of scalar multipole solitons in the optical media with
 thermal nonlocal nonlinearity has been
 reported \cite{SegevExperiment}. It is particularly remarkable
 that all multisolitons investigated in Ref. \cite{SegevExperiment}
 are found to be unstable, but with a large parameters range where
  the instability is weak, so that their experimental observation is possible.
 In this Letter, using direct 2D simulations for a scalar model with thermal
 nonlocal nonlinearity, we find a
class of radially asymmetric nonrotating two-dimensional soliton
solutions and show that, at some input power, the dipole-mode
scalar solutions are stable, while the tripoles and quadrupoles
are always unstable, but can survive over quite considerable
distances. The stability window for dipolar multisolitons is
found. We find also rotating dipole and quadrupole solutions
(azimuthons) which turn out to be stable if the beam power exceeds
some threshold value.


The basic system of equations, written in appropriate
dimensionless variables, is
\begin{equation}
 \label{NLS}
 i\frac{\partial \Psi}{\partial z}+\Delta_\perp\Psi +\theta\Psi=0,
\end{equation}
\begin{equation}
\label{Theta}
 \alpha^{2}\theta-\Delta_\perp\theta=|\Psi|^2,
\end{equation}
where $\Delta_\perp=\partial^2/\partial x^2+\partial^2/\partial
y^2$ is the transverse Laplacian.  Equations (\ref{NLS}) and
(\ref{Theta}) describe the propagation in $z$-direction of the
electric field envelope $\Psi(x,y,z)$ coupled to the temperature
perturbation $\theta(x,y,z)$ in a plasma \cite{LitvakSJPlPhys75}.
The identical model describes the wave field $\Psi$ and spatial
distribution of the molecular director $\theta$ in nematic liquid
crystals~\cite{ContiPRL03}.

The parameter $\alpha$ stands for the degree of nonlocality of the
nonlinear response. In the limit $\alpha^{2}\gg 1$, Eqs.
(\ref{NLS}) and (\ref{Theta}) reduce to the usual nonlinear
Schr\"{o}dinger equation; the opposite case $\alpha^{2}\ll 1$
corresponds to a strongly nonlocal regime. For plasmas parameter
$\alpha$ characterizes the relative energy that electron with mass
$m$ delivers to a heavy particle with mass $M$ during single
collision ($\alpha^2\approx 2m/M$). Therefore, the thermal
self-focusing in a plasma occurs in a strongly nonlocal regime.
The model used in Ref. \cite{Segev} to describe the propagation of
the wave beam in optical media with thermal nonlinearity
corresponds to the specific nonlocal limit $\alpha\to 0$. In
nematic liquid crystals, the degree of nonlocality can be
modulated by changing the pretilt angle $\theta_0$ of molecules
through bias voltage $V$ \cite{HuDipole}. As $V$ increases, the
degree of nonlocality decreases, so that the degree of nonlocality
can be tuned from local regime to a strongly nonlocal one.

Equations (\ref{NLS}) and (\ref{Theta}) can be rewritten as a
single integro-differential equation
\begin{equation}
\label{gen}
 i\frac{\partial \Psi}{\partial z}+\Delta_\perp\Psi +
 \Psi\int
 R(|\mathbf{r}-\mathbf{r}'|)|\Psi(\mathbf{r}')|^{2}\,d\mathbf{r}'=0,
\end{equation}
with the kernel $R(\xi)=K_{0}(\alpha\xi)/(2\pi)$, where $K_{0}(z)$
is the modified Bessel function of the second kind of order zero.
It is seen, that the nonlinearity in Eq. (\ref{gen}) has
essentially nonlocal character.

\begin{figure}[e]
\begin{center}\includegraphics[width=5in]{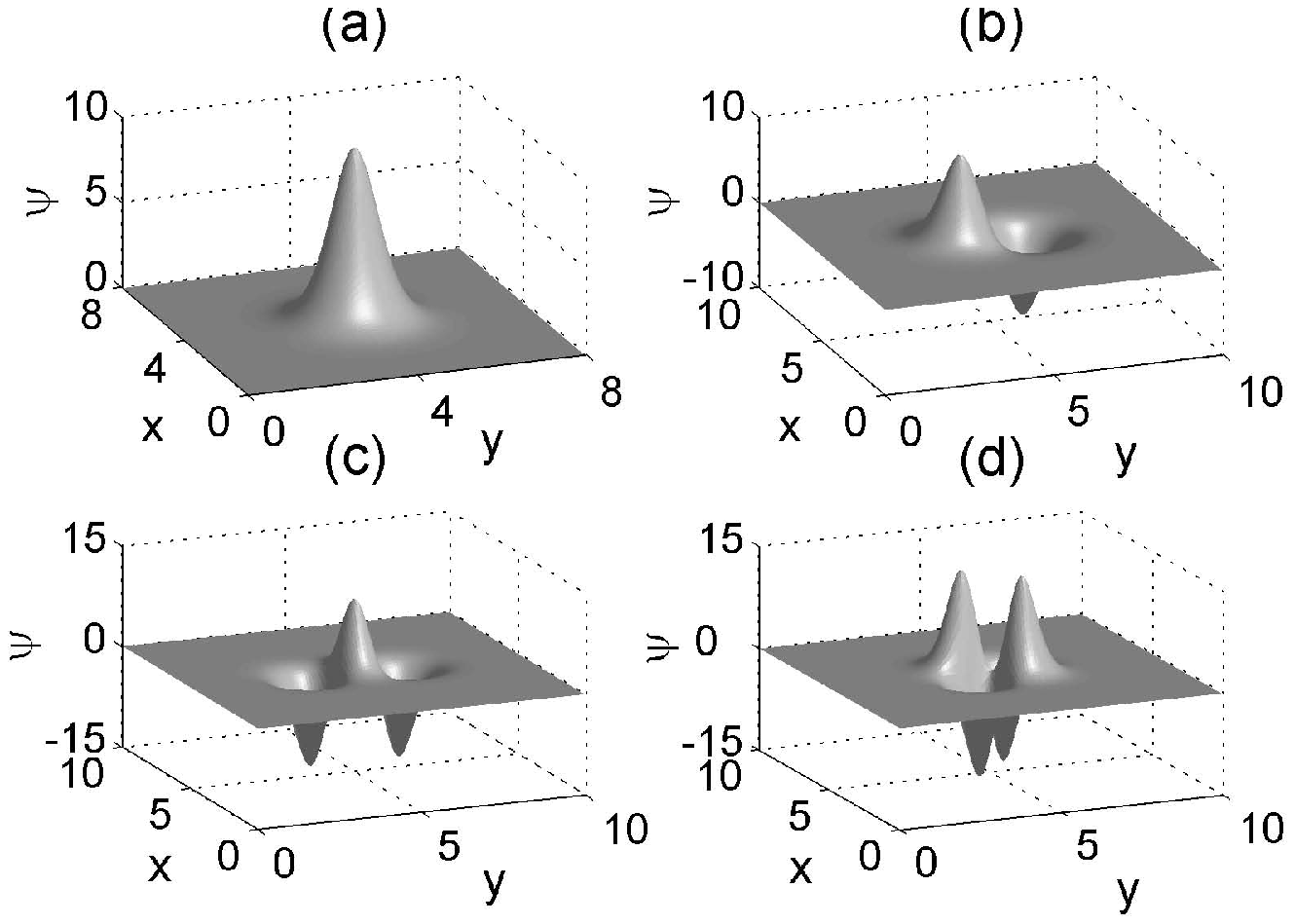}\end{center}
\caption{\label{NumericalSteadyState} Numerically found stationary
localized nonrotating solutions of Eqs. (\ref{eq1}) and
(\ref{eq2}): (a) monopole with $\lambda=10$; (b) dipole with
$\lambda=10$; (c) tripole with $\lambda=15$; (d) quadrupole with
$\lambda=20$.}
\end{figure}

\begin{figure}[e]
\begin{center}\includegraphics[width=5in]{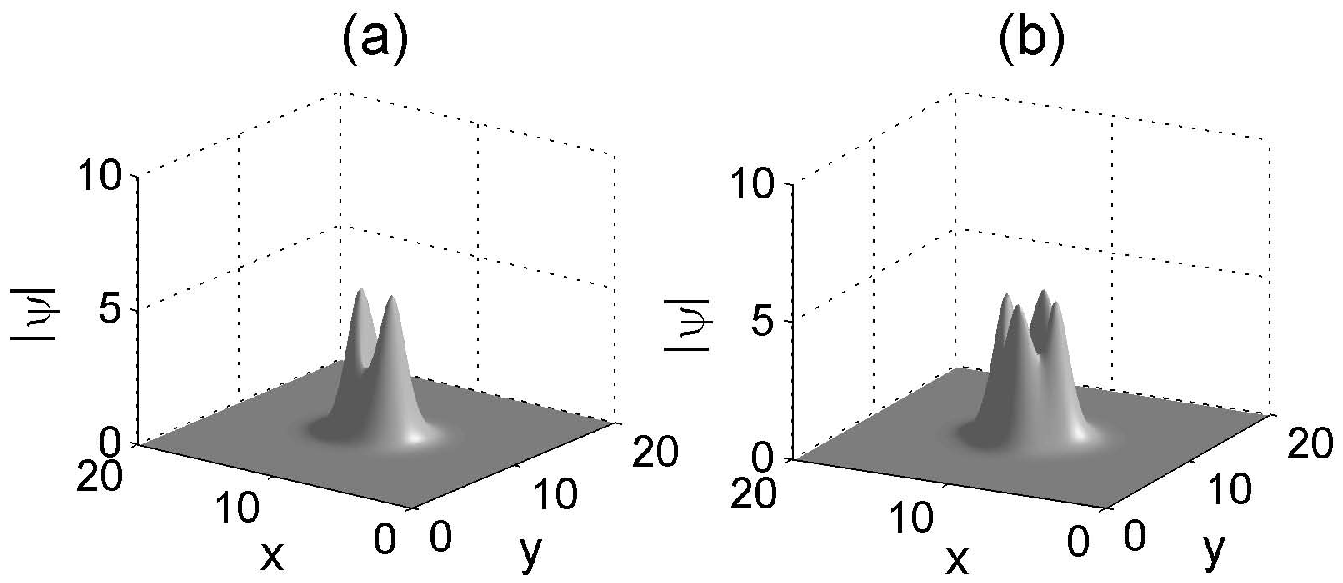}\end{center}
\caption{\label{Azimut} Two examples of the azimuthons with
$\lambda=4.5$. Intensity $|\psi|$ is shown in the $(x,y)$ plane.
a) solution with two intensity peaks; b) solution with four
intensity peaks.}
\end{figure}

We look for stationary solutions of Eqs. (\ref{NLS}) and
(\ref{Theta}) in the form $\Psi(x,y,z)=\psi(x,y)\exp(i\lambda z)$,
so that $\psi(x,y)$ obeys the equations
\begin{equation}
 \label{eq1}
 -\lambda\psi+\Delta_\perp\psi +\theta\psi=0,
\end{equation}
\begin{equation}
\label{eq2} \alpha^{2} \theta-\Delta_\perp\theta=|\psi|^2,
\end{equation}
 and we do not assume the radial symmetry of $\psi(x,y)$. For the
 numerical modeling we use the rescaled variables $\psi\rightarrow\psi/\alpha$,
  $\theta\rightarrow\theta/\alpha^{2}$, $z\rightarrow z\alpha^{2}$,
   $(x,y)\rightarrow (x,y)\alpha$, $\lambda\rightarrow\lambda/\alpha^{2}$, so that
 $\alpha=1$ in Eqs. (\ref{eq1}) and (\ref{eq2}). Note,
  a strongly nonlocal regime ($\alpha\ll 1$) corresponds to
  large values of the scaled propagation constant $\lambda$.
Imposing periodic boundary conditions on Cartesian grid and
choosing an appropriate initial guess we have found a class of
radially asymmetric multipole localized solutions by using the
relaxation technique similar to one described in Ref.
\cite{Petviashvili86}. The real (or containing only a constant
complex factor) function $\psi(x,y)$ corresponds to nonrotating
solitary structures. Examples of such nonrotating multipole
solitons, namely, a dipole, a tripole, and a quadrupole are
presented in Fig.~\ref{NumericalSteadyState}. The nonrotating
multipoles consist of several fundamental solitons (monopoles)
with opposite phases. The complex function $\psi(x,y)$ with a
spatially modulated phase corresponds to rotating structures
(azimuthons) with nonzero angular momentum. In Fig.~\ref{Azimut}
we demonstrate two examples of the azimuthons for the nonlocal
model described by Eqs. (\ref{NLS}) and (\ref{Theta}).

To gain the better insight into the properties of the stationary
nonrotating multisolitons we have performed the analytical
variational analysis. A stationary nonrotating multisoliton can be
described by the trial function of the form:
\begin{equation}\label{ansatz}
\Psi(x,y,z)=hf(x,y)e^{i\lambda z},
\end{equation}
where the shape of the dipolar multisoliton can be approximated by
the antisymmetric superposition of two gaussian functions:
$$f(\xi,\eta)=e^{-\frac{1}{2a^2}\left\{y^2+(x-d/2)^2\right\}}-e^{-\frac{1}{2a^2}\left\{y^2+(x+d/2)^2\right\}},$$
where $a$ and $d$ characterize the radius of each monopoles and
the distance between them, respectively. For the bound state of
$S$ identical single solitons with the centers at $(x_s,y_s)$,
where $s=1\dots S$, we have used the trial function of the form:
$$
f(\xi,\eta)=\sum_{s=1}^Sg_s
e^{-\frac{1}{2}(\xi-\xi_s)^2-\frac{1}{2}(\eta-\eta_s)^2},$$ where
$(\xi,\eta)=(x,y)/a$, $g_s=\pm 1$ gives the phase of the $s$-th
soliton.

As known, the steady-state corresponds to the stationary point of
the Hamiltonian
\begin{equation}  \label{Hamilt2D}
 H = \int\left\{| \nabla_\perp \Psi|^2-\frac12\theta |\Psi|^2
 \right\}d^2\textbf{r}
\end{equation}
at the fixed number of quanta $N$
\begin{equation}  \label{NumbQuant}
N = \int|\Psi|^2d^2\textbf{r}.
\end{equation}
In the variational approach, Hamiltonian $H(a,b)$ is the function
of two variational parameters $a$ and $b=d/a$, where $a$
characterizes the width of each monopole and $d$ is the distance
between two neighboring out-of-phase solitons. The third parameter
of the trial function (\ref{ansatz}), the amplitude $h$, has been
 eliminated in terms of $a$, $b$, and $N$ using the normalization
condition (\ref{NumbQuant}).

\begin{figure}[e]
\begin{center}\includegraphics[width=\textwidth]{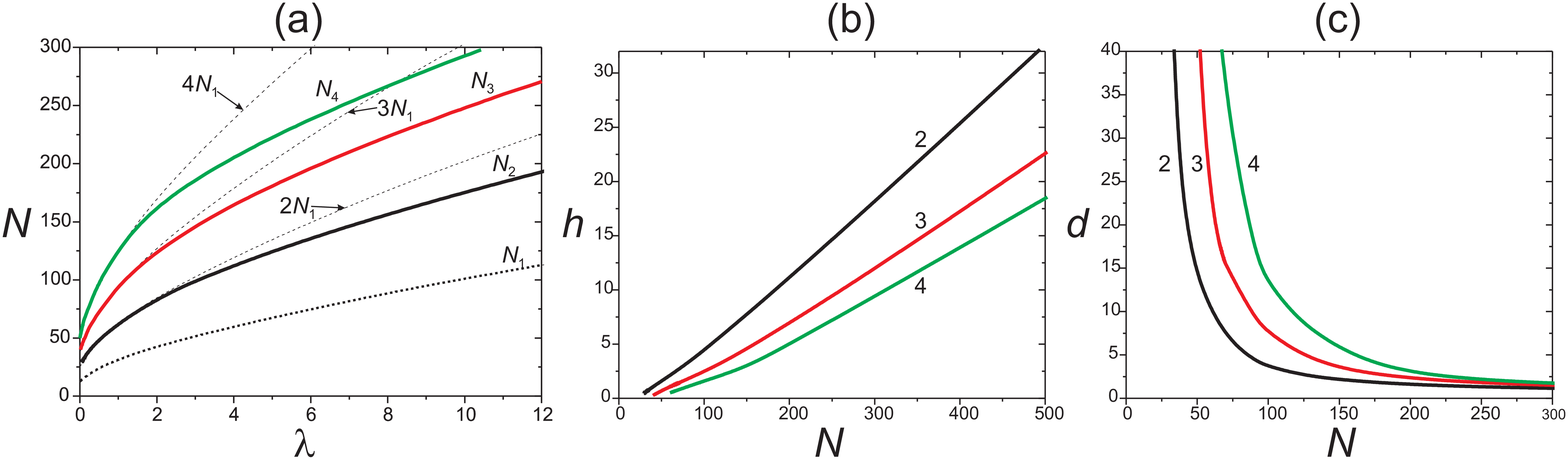}\end{center}
\caption{\label{Variat} (a) Number of quanta $N$ vs propagation
constant $\lambda$. Solid curves for the bound states of $S$
out-of-phase solitons, dashed curves for $SN_1$ (variational
results). The integers near the curves indicate $S$, the number of
solitons.}
\end{figure}

Results of the variational analysis are found to be in very good
agreement with our numerical simulations for stationary
nonrotating multisolitons. Figure \ref{Variat} (a) shows the beam
power $N$ versus the propagation constant $\lambda$ for monopoles,
dipoles, tripoles, and quadrupoles. The threshold power for
existence of the stationary bound state of $S$ solitons is very
close to the threshold for formation of $S$ single unbound
fundamental solitons. Moreover, the "binding energy" $N_S-SN_1$
(where $N_1$ is the power for single fundamental soliton) is very
small for $\lambda<1.9$, thus these multisolitons should readily
decay into inbound solitons. As is seen from Fig. \ref{Variat}
(c), the distance between neighboring out-of-phase solitons $d$
increases dramatically when $\lambda<1.9$ which lead to decreasing
of interaction between monopoles.

We next addressed the stability of these multipole solutions and
study the evolution (propagation) of the dipoles, tripoles,
quadrupoles, and azimuthons in the presence of small initial
perturbations. We have undertaken extensive numerical modeling of
Eqs. (\ref{NLS}) and (\ref{Theta}) initialized with our
numerically computed multipole solutions with added gaussian
noise. The initial condition was taken in the form
$\psi(x,y)[1+\varepsilon \Phi(x,y)]$, where $\psi(x,y)$ is the
numerically calculated exact multipole solution, $\Phi(x,y)$ is
the white gaussian noise with variance $\sigma^{2}=1$ and the
parameter of perturbation $\varepsilon=0.005 \div 0.1$. Spatial
discretization was based on the pseudospectral method and
"temporal" $z$-discretization included the split-step scheme.

\begin{figure}[e]
\begin{center}\includegraphics[width=5in]{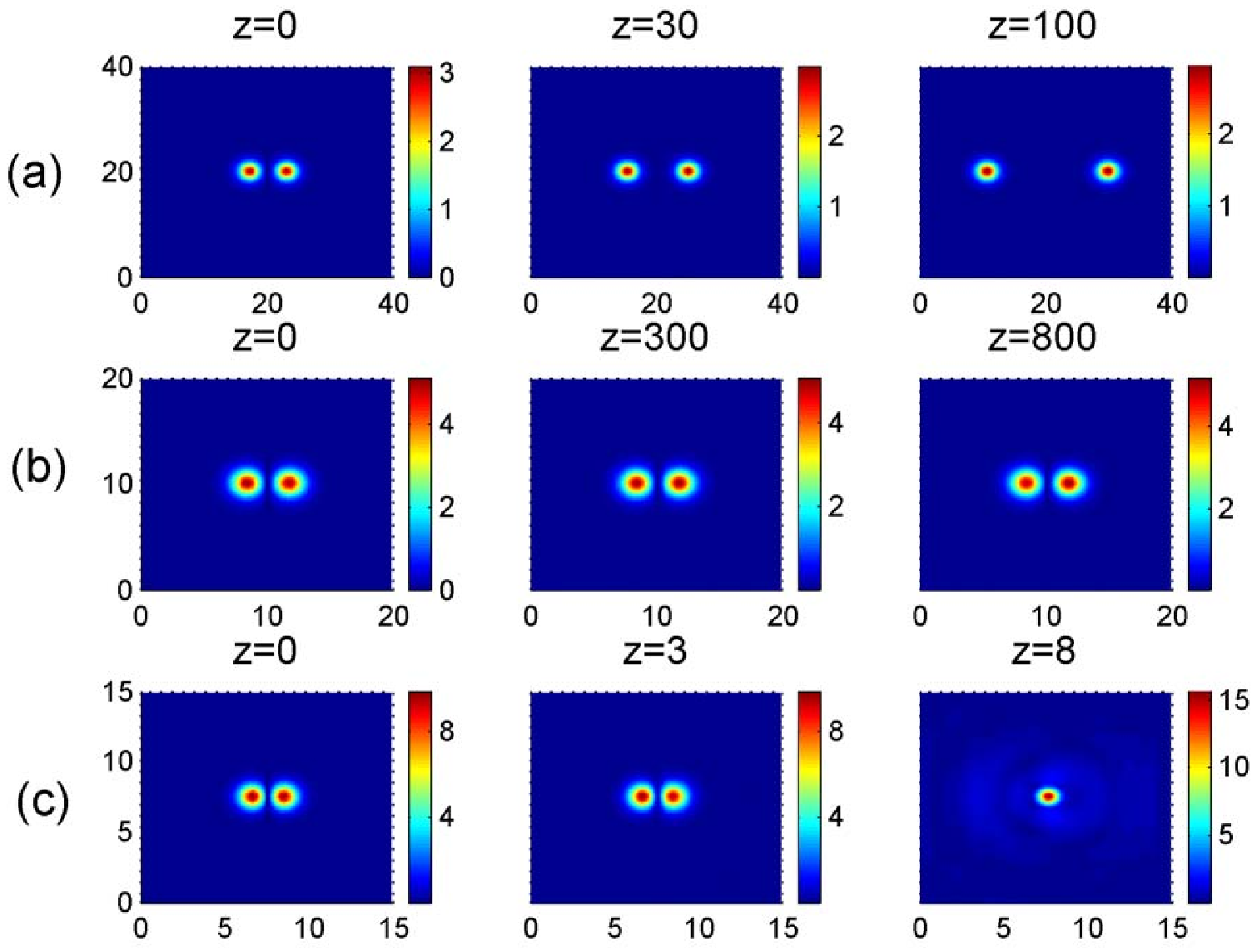}\end{center}
\caption{\label{Dynam1} (a) Splitting of the dipole with
$\lambda=1.5$ into two monopoles; (b) stable propagation of the
dipole with $\lambda=3.5$; (c) unstable dipole with $\lambda=10.$
}
\end{figure}

 Depending
on the parameter $\lambda$, we observed three different regimes of
the nonrotating dipole propagation, which are presented in
Fig.~\ref{Dynam1} (for $\varepsilon=0.02$). The first regime
corresponds to the region $\lambda<\lambda_{cr}$, and we found
$\lambda_{cr}\sim 1.9$. If $\lambda<\lambda_{cr}$, the initial
dipole splits in two monopoles which move in the opposite
directions without changing their shape and without radiation,
i.~e. the monopoles just go away at infinity. This type of the
evolution is shown in Fig.~\ref{Dynam1}(a). The splitting of the
dipole into two monopoles can be easily understood. It is seen
from Fig. \ref{Dynam1} that the bound energy $\delta
N=N_{dip}-2N_{mon}$ in the dipole tends to almost zero as
$\lambda$ approaches $\lambda_{cr}\sim 1.9$. This explains why the
dipole with $\lambda\leq \lambda_{cr}$ can be easily (i. e. under
the action of extremely small initial perturbations) split into
two monopole-type solitons. A similar behavior, i.e. the decay of
the initial dipole into two stable moving monopoles below some
critical value of $\lambda$, was observed for the model with a
Gaussian response function $R(\xi)$ in Eq. (\ref{gen}) \cite{We}.

\begin{figure}[e]
\begin{center}\includegraphics[width=4in]{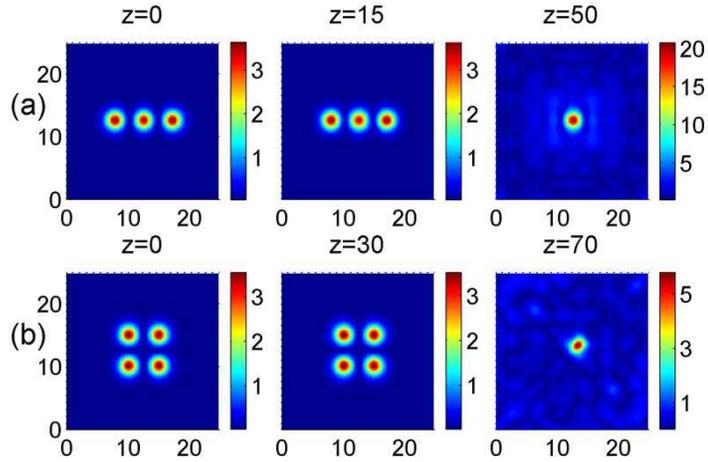}\end{center}
\caption{\label{Dynam2}Typical unstable evolution (at $\lambda=2$)
of (a) tripole; (b) quadrupole.}
\end{figure}

\begin{figure}[e]
\begin{center}\includegraphics[width=4in]{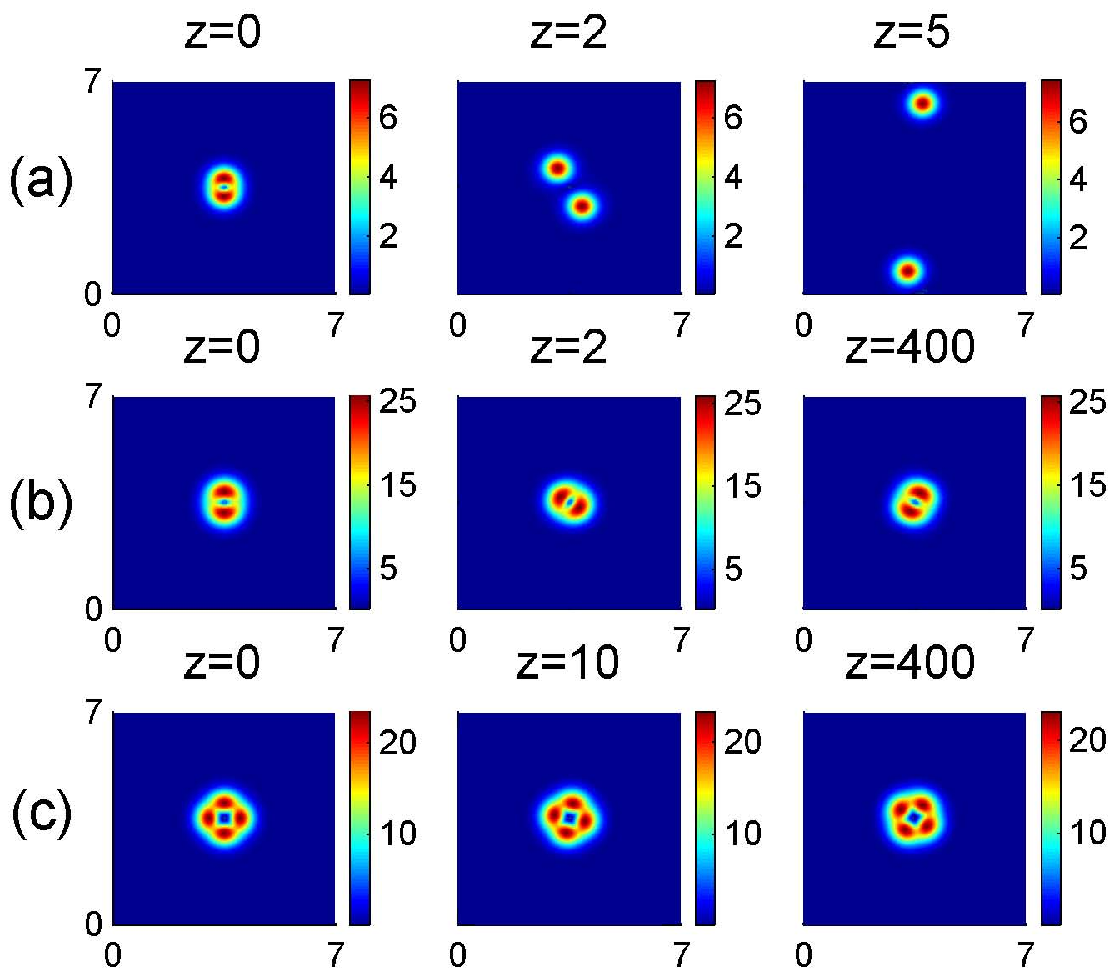}\end{center}
\caption{\label{Dynam3} (a) Splitting of the azimuthon with two
intensity peaks and $\lambda=10$; (b) stable propagation of the
azimuthon with two intensity peaks and $\lambda=80$; (c) stable
propagation of the azimuthon with four intensity peaks and
$\lambda=80$.}
\end{figure}

The second regime corresponds to the region
$\lambda_{cr}<\lambda<\lambda_{th}$, where $\lambda_{th}\sim 4$.
The numerical simulations clearly show that in this range of the
parameter $\lambda$ the dipoles are stable with respect to initial
noisy perturbations. If the parameter of perturbation
$\varepsilon$ is not too large, the dipoles survive over huge
distances ($z>3000$). The stable propagation of the dipole is
illustrated in Figs.~\ref{Dynam1}(b) (for $\lambda=3.5$ and
$\varepsilon=0.02$).

The further (after $\lambda_{th}\sim 4$) increasing the parameter
$\lambda$ sharply shortens the propagation distances at which the
dipole survives, and, the dipoles with $\lambda>\lambda_{th}$ are
unstable. The typical decay of the unstable dipole above the
threshold value $\lambda_{th}$ of the rescaled propagation
constant is shown in Figs.~\ref{Dynam1}(c). Thus, the stable
dipoles exist only within a finite, rather narrow range of the
propagation constants $\lambda$.

Figure~\ref{Dynam2} illustrates the propagation of the tripole and
quadrupole for $\lambda=2$, i.e. in the region, where the dipole
is stable. Generally speaking, the tripoles and quadrupoles turn
out to be unstable, but for $\lambda_{cr}<\lambda<\lambda_{th}$
they can survive on the quite considerable (compared to the
characteristic diffraction length) distances and, thus, can be
experimentally observed. Tripoles and quadrupoles with
$\lambda<\lambda_{cr}$ decay into three and four monopoles
respectively.

We observed two different scenarios for the azimuthon propagation.
If $\lambda$ (i. e. the beam power) is small enough
($\lambda<15$), the azimuthons are unstable and split into
fundamental solitons (monopoles) which move away from each other.
An example of such splitting for the azimuthon with two intensity
peaks and $\lambda=10$ is presented in Fig.~\ref{Dynam3}(a). For
sufficiently large $\lambda$ ($\lambda > 15$), the azimuthons turn
out to be stable. Stable propagation of the azimuthons with two
(rotating dipole) and four (rotating quadrupole) intensity peaks
and $\lambda=80$ is shown in Figs.~\ref{Dynam3}(b),(c). Note that
stable rotating dipoles were also observed in Ref.~\cite{Skupin}.
Numerically estimated rotational velocity of the stable azimuthon
in Fig.~\ref{Dynam3}(b) is $\omega=1.15$ so that it survives over
hundreds rotational periods.

The dynamics of the nonrotating dipole in our model Eqs.
(\ref{NLS}) and (\ref{Theta}) is in sharp contrast to the dipole
propagation in the model with a Gaussian response function
$R(\xi)$ in Eq. (\ref{gen}), where the stable nonrotating dipoles
were observed for all $\lambda>\lambda_{th}$, where $\lambda_{th}$
is some threshold value \cite{Kivshar3,We}. A qualitatively
different behavior of the dipoles in these models seems to be
related to the regularity properties of the functions $R(\xi)$ in
Eq. (\ref{gen}) -- for the model described by Eqs. (\ref{NLS}) and
(\ref{Theta}) the function $R(\xi)$ has a singularity at zero. The
strong dependence of a stability criteria for multisoliton
solutions on the regularity properties of the kernel $R$ in Eq.
(\ref{gen}) was discussed also in Ref. \cite{We,Skupin}. The
corresponding theoretical problem seems to be intriguing and
highly nontrivial.


In conclusion, we have studied the bound states of the
out-of-phase nonrotating two-dimensional solitons and rotating
multisolitons in nonlocal nonlinear media. We have demonstrated
that stationary nonrotating dipolar, tripolar, quadrupolar
multisolitons and rotating multisolitons (azimuthons) may exist in
media with thermal nonlocal nonlinearity, if the beam power is
above some critical value. We have investigated stability of the
multisolitons using direct numerical simulations. The nonrotating
tripoles and quadrupoles are found to be unstable with respect to
decay into fundamental solitons or fusion into single soliton,
depending on the beam power. At the same time, we have observed
robust propagation of the azimuthons with sufficiently high energy
and a dipolar multisoliton with moderate energy. Thus, our
theoretical predictions open the prospects for the experimental
observations of the stable azimuthons and the bound states of two
out-of-phase wave beams in media with thermal self-focusing
nonlocal nonlinearity.

\end{document}